\documentclass[12pt]{article} 
\usepackage{graphicx}
\usepackage{curvesls}
\usepackage{epsfig}
\def\mathbf{\vec}

\begin{document}
\centerline {\LARGE Complex wave dynamics in microlasers}
\vspace{.3cm}
\centerline {\LARGE and nonlocal maps.}
\vspace{1cm}
\centerline {\large A.Yu.Okulov}
\vspace{.5cm}
\centerline {\it P.N.Lebedev Physical Institute of Russian Academy of Sciences}
\centerline {\it Leninsky  prospect  53, 119991 Moscow, Russia}
\centerline {\it   okulov@sci.lebedev.ru}
\centerline {\it and}
\centerline {\it Department of Physics, University de Coimbra,}
\centerline {\it  3004-516 Coimbra, Portugal}
\vspace{1cm}
\begin{abstract}
The optical  implementation of complex network is considered for laser cavity formed by 
active mirrors deposited on the surface of plane-parallel slice, ellipsoidal microball , whose eccentricity 
is close to 0.5 and equivalent optical circuit, containing 
amplifying and absorbing media in Fourier-conjugated planes. In this types of microlaser's
the self-imaging cavity provides efficient diffractive interconnections between different points 
of the mirrors. The interconnections could be local, as in plane-parallel geometry, global as in confocal cavity or 
inhomogeneous one with definite subset of nodes, having much more interconnections than their neighbours.
 We show how the spatiotemporal dynamics of this nonlinear resonator is 
controlled by the geometry of gain-losses distribution making  possible the analog modeling 
of the several  generic  nonlinear evolution equations. 
\end{abstract}
\centerline {\large PACS number(s:)      05.45-a., 42.60.-v, 45.70.Qj}
\vspace{1cm}
\section{Introduction.}
The close analogy between spatiotemporal patterns in optics and hydrodynamics is being the subject of 
relentless interest within recent decades \cite{Oraevsky:1964,Lorenz:1963,Newell:1990,Okulov:1988}.  Inspite of a substantial difference 
between Maxwell-Bloch and Navier - Stockes equations,
 the definite similarity of dynamical behaviour have been  already established
 \cite{Lugiato:1991,Staliunas:1993}. It is well known that 
both set of equations are reduced with the some degree of accuracy to the Ginzburg-Landau (GLE)\cite{Newell:1990}  or to the 
Swift-Hohenberg (SHE) \cite{Staliunas:1993} equations. The last equation contains additional fourth-order dispersion terms, which arrests 
collapse, inherent to 2D GLE and stabilizes the spatial solitons  \cite{Staliunas:1993}. The competition between localized 
structures \cite{Rosanov:2003, Lugiato:1998}, 
stripes, lattices and chaos is shown to be highly sensitive also to boundary conditions, which could be controlled by 
the proper choice of the Fourier spatial filter \cite{Firth:1998} placed inside the cavity. In contrast to the case of spatial filtering by finite gain 
linewidth \cite{Staliunas:1993}, the boundary 
 conditions itself, namely the spatial layout of the gain elements, diaphragms and mirrors provides the possibility to 
control the spatiotemporal dynamics \cite{Okulov:1988,Firth:1998}. In this context the situation in laser cavity is similar 
to the complex networks physics, where 
spatial interconnections define both dynamical and statistical properties of network \cite{Barabashi:2002}. 

It is shown in a set of interesting cases that conventional set of Maxwell-Bloch 
equations for class-A laser could be reduced by split-step Green function technigue to the integral equation including the boundary 
conditions in explicit form \cite{Okulov:1988,Okulov:2000} :
\begin{equation}
\label{nonlmap}
\ E_{n+1}(\vec{r} )= \int^\infty_{-\infty}\int^\infty_{-\infty}\ K(\vec{r}-\vec{r'}) f(E_{n}(\vec{r'})) d^2 \vec{r'}
\end{equation}
This equation has a form of the product of two operators in the form of convolution: 
\begin{equation}
\label{nonlmap1}
\ E_{n+1}(\vec{r} )= \hat{K} f(E_{n}(\vec{r})) 
\end{equation}
where $ {E_n}$ is amplitude of electromagnetic field, $  \hat{K}$ is linear nonlocal operator  (propagator of the cavity),  $  {f}$ is nonlinear local operator obtained under thin slice approximation
\cite{Okulov:1988}. For example, the kernel  $  {K}$ for  the  perfectly  plane-parallel Fabry-Perot cavity of microchip laser 
has the form \cite{Weinstein:1969,Okulov:1990}: 
\begin{equation}
\label{kernelfabry}
\ K(\vec{r}-\vec{r'} )= \frac{ik}{2 \pi z}     exp (ik(\vec{r}-\vec{r'})^2 / 2 z) 
\end{equation}
the transfer function  $ { f[E_{n}(\vec{r}), G(\vec{r}) ] }$ explicitly contains the gain distribution $ {G(\vec{r})}$ for this  microchip laser cavity configuration 
(fig.1) \cite{Okulov:2000,Okulov:1990,Chen:2001,Okulov:2002}:

\begin{figure}
\begin{picture}(100,100)(-100,-20)
\thicklines
\put(10,50){\framebox(20,100)}
\put(-50,115){$y$}
\put(-63,33){$x$}
\put(-16,57){$z$}
\put(-50,60){\vector(1,0){30}} 
\put(-50,60){\vector(0,1){50}} 
\put(-50,60){\vector(-1,-2){10}} 

\put(-20,120){\vector(1,0){30}} 
\put(-20,110){\vector(1,0){30}} 
\put(-20,100){\vector(1,0){30}} 
\put(-20,90){\vector(1,0){30}} 
\put(-20,80){\vector(1,0){30}}
\put(100,50){\line(0,1){100}} 
\put(13,50){\line(0,1){100}} 
\put(88,30){\curve(1,20,6,70,1,120)}
\put(85,30){\curve(1,20,6,70,1,120)}
\put(10,100){\oval(15,35)[r]}
\put(10,100){\oval(20,40)[r]}
\put(10,100){\oval(25,45)[r]}
\put(10,100){\oval(30,50)[r]}
\put(86,50){\line(1,0){15}}  
\put(86,150){\line(1,0){15}}  
\put(105,120){\vector(1,0){30}} 
\put(105,100){\vector(1,0){30}}
\put(105,80){\vector(1,0){30}}
\put(-10,20){$Nd:YAG$}
\put(-10,7){$slice$}
\put(-30,133){$ {\lambda}$}
\put(113,131){$ {\lambda}$}
\put(-25,125){$ pump $}
\put(120,125){$ osc $}
\put(-10,35){$ d\approx500  {\mu}m$}
\put(59,125){$ G({\vec{r}}) $}
\put(57,125){\vector(-2,-1){30}} 
\put(39,95){$ L_{r} \approx  D $}
\put(39,80){$ \approx 10^3 {\mu}m $}
\put(80,167){$Output Mirror$}
\put(105,155){$R\approx 0.97$}
\put(3,155){$R\approx 0.999$}
\put(209,131){$ Intensity $}
\put(223,120){$ 2D $}
\put(216,110){$ vortex $}
\put(216,100){$ array $}

\end{picture}
\caption{High-$ Q $ microchip solid-state laser cavity\cite{Chen:2001}
. The length $ z=L_{r} $ is close to beam diameter $ D $. The transversal ${\vec{r}}=(x,y)$ intensity distribution in 2D vortex array inline.}
\label{fig1}
\end{figure}
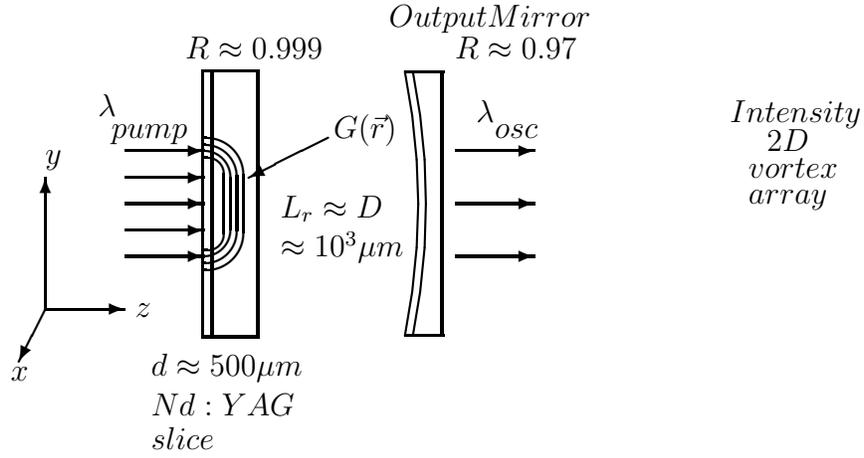
\begin{figure}
\begin{picture}(100,100)(-100,-20)
\thicklines
\put(-50,115){$z$}
\put(-63,33){$x$}
\put(-16,57){$y$}
\put(-50,60){\vector(1,0){30}} 
\put(-50,60){\vector(0,1){50}}
\put(-50,60){\vector(-1,-2){10}}
\put(120,110){$ G({\vec{r"}}) $}
\put(118,109){\vector(-3,-1){30}}
\put(150,115){\vector(4,-1){55}}

\put(80,27){\vector(1,2){32}} 

\put(80,63){\vector(0,1){32}} 

\put(100,60){$ R_c $}
\put(69,41){$ F $}

\put(84,72){$ F $}

\put(80,61){\vector(0,-1){32}} 
\put(120,11){$ f(E_{n}({\vec{r}})) $}
\put(118,15){\vector(-3,1){30}}
\put(168,12){\vector(4,1){40}}
\put(60,7){\curve(60,84,20,93,-20,84)}
\put(60,7){\curve(90,70,20,90,-50,70)}
\put(60,7){\curve(90,40,20,20,-50,40)}
\put(60,7){\curve(90,70,100,55,90,40)}
\put(60,7){\curve(-50,70,-60,55,-50,40)}
\put(60,7){\curve(60,26,20,17,-20,26)}

\put(-16,0){$a)$}
\put(270,0){$b)$}

\put(210,25){\framebox(40,5)}
\put(210,100){\framebox(40,5)}
\put(210,10){\framebox(40,3)}
\put(210,160){\framebox(40,3)}

\put(210,11){\line(1,0){40}} 

\put(210,162){\line(1,0){40}}
\put(130,-5){$ Saturable $}
\put(130,-15){$ absorber $}

\put(210,-5){$ Mirror $}
\put(210,169){$ Mirror $}
\put(120,129){$ Gain $ $slice $}

\put(220,114){$ F $}

\put(220,76){$ F $}

\put(230,136){\vector(0,-1){32}} 

\put(230,68){\vector(0,1){32}} 

\put(210,63){\curve(0,0,20,5,40,0)}

\put(210,63){\curve(0,0,20,-5,40,0)}

\put(210,141){\curve(0,0,20,5,40,0)}
\put(210,141){\curve(0,0,20,-5,40,0)}

\end{picture}
\caption{a)Quasiconfocal cavity formed by active mirrors on the surface of the ellipsoidal microball\cite{Okulov:2000}
. The focal length of active mirrors $ F $ is a half of raduis of curvature  $ R_c $ . b) Quasiconfocal cavity with two 
lenses and two plane-parallel nonlinear elements: $ G({\vec{r"}}) $ and $ f(E_{n}({\vec{r}})) $ \cite{Staliunas:1998}. }
\label{fig2}
\end{figure}
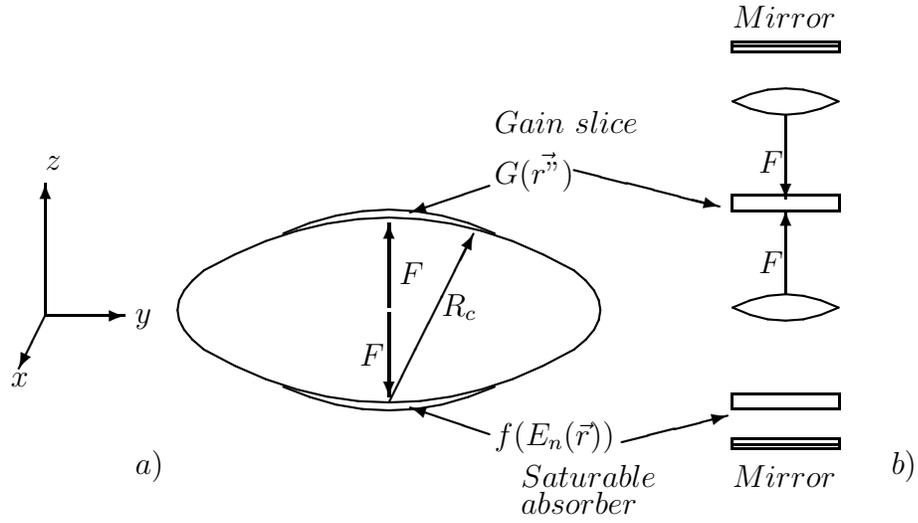
On the other hand, for the perfectly confocal cavity  (fig.2), the kernel $ K $   is Fourier transform of diaphragm transmission. 
For example, when confocal cavity is formed by two mirrors, located in Fourier - conjugated planes, one being active, 
the kernel  $ K $ becomes the Fourier transform of gain distribution $ G $  \cite{Okulov:1988,Okulov:2000}:
\begin{equation}
\label{kernelconf}
\ K(\vec{r}-\vec{r'} )= \int^\infty_{-\infty}\int^\infty_{-\infty}  \ exp(ik(\vec{r}-\vec{r'}) \vec{r''} / 2 z)G ( \vec{r''} ) d^2 \vec{r''}
\end{equation}
The confocal cavity could be formed by two concave mirrors with the common focus \cite{Okulov:2000}, four mirrors in 8-F ring laser geometry
 \cite{Okulov:1988} or inside the ellipsoidal microball (fig.2a). The equivalent optical circuit formed by two confocal lenses and two planar, gain 
and absorber layers, had been used for excitation and detection of the spatial solitons (fig.2b) \cite{Staliunas:1998}.  
In the plane-parallel geometry (fig.1) the interconnections occur mainly 
in between adjacent points of the mirror (neighbour-neibourhood interaction), while in confocal geometry (fig.2) the interconnections 
are global: each point interacts with the others through gain layer, located in "far field" . 
It is clear enough  that the intermediate situation in 
between plane - parallel (fig.1) and confocal (fig.2) configurations, is described by equation (\ref{nonlmap}) as well, 
provided the kernel $ K $  contains terms corresponding to both difraction in free space and absorbance at the diafragm (''diffusion''). 
This intermediate situation with possibility of creating the selected "nodes" having much more interconnections than neighbours 
is interested from the point of view of complex networks theory \cite{Barabashi:2002}.

We will see that initial master equation  (\ref{nonlmap}) having the form of nonlinear nonlocal map, 
the nonlocality being induced by the explicit inclusion of the boundary conditions \cite{Okulov:1988}, 
is a quite general form of  intracavity laser dynamics.
Originally this model had been used for modeling of travelling 
wave tube microwave oscillator wih delayed feedback and some other relevant systems \cite{Pikovsky:1986}, the scaling relations for a weak 
chaos had been established. Afterwards the same model, called "continuous coupled map" (or "continuous coupled map lattice"), 
had also been used for description of 
the pattern formation in vibrating granular layer  \cite{Ott:2001}. 
Our goal here is to show that nonlinear dynamics described 
by primary master equation in the form of nonlocal map (\ref{nonlmap}) contains the different types of nonlinear behaviour, 
intrinsic to a wide set of several nonlinear 
evolution equations. The dynamics inherent to Burgers, nonlinear Shrodinger, Ginzburg-Landau or Swift-Hohenberg equations will be 
released by the proper choice of gain-losses distribution.  
\section{High gain limit. }
In the high gain limit, when the field changes in nonlinear slice are assumed to  be arbitrarily large 
compared to diffraction effects, we have : 
\begin{eqnarray}
\label{nonloc2}
\ E_{n+1}(\vec{r} )
 & =&  f (E_{n}(\vec{r}))  + \alpha_1 \frac {\partial f}{\partial E} \frac {\partial E_n}{\partial x} + 
	{\alpha_2}\left[\frac {\partial f}{\partial E}\frac {\partial^2 E_n}{\partial x^2}+
\frac {\partial^2 f}{\partial E^2}\frac {\partial E_n}{\partial x}\ \right] + \nonumber \\
&+& {\alpha_3}\left[\frac {\partial f}{\partial E}\frac {\partial^3 E_n}{\partial x^3}+
		3\frac {\partial^2 f}{\partial E^2}\frac {\partial E_n}{\partial x}\frac {\partial^2 E_n}{\partial x^2}
	+\frac {\partial f^3}{\partial E^3}\left(\frac {\partial  E_n}{\partial x}\right)^3\ \right] + ... 
\end{eqnarray}
where   
\begin{equation}
\label{moment}
\alpha^m = \frac{1}{m!}\int^\infty_{-\infty}\ (x-x')^m K(x-x')  dx'
\end{equation}
are the moments of  kernel  $ K $, the one transverse dimension  $ x $ is considered here and further for simplification of expressions.
Now we ought to take into account the components of the gain function $ f $ having different orders of magnitude : 
\begin{equation}
\label{nonlin}
f (E_{n}(\vec{r}))= G d E_{n} \left[ 1 - {\beta}_s |E_{n}|^2\right]
\end{equation}
Obviously, this form of the gain function corresponds to the two-level medium with weak saturation. 
When   $  {\alpha_2},{\alpha_3} = 0$  we have the wave equation with nonlinear source $ f (E_{n}(\vec{r}))$:  
\begin{equation}
\label{shock}
\ E_{n+1}(\vec{r} ) = f (E_{n}(\vec{r}))  +  \left[\alpha_1 \frac {\partial f}{\partial E}\right] \frac {\partial E_n}{\partial x} 
\end{equation}
The transition from "discrete time"  $n$ with step  $ \delta{t}=\frac{L_{r}}{2c}$ ($c$ - is the speed of light ) to continuous time $t$ 
is straightforward and we can get in the so-called "mean-field approximation"
\cite{Lugiato:1991,Staliunas:1993,Suchkov:1965} the equation for the order parameter: 
\begin{equation}
\label{shock1}
\frac {\partial E(\vec{r},t)}{\partial t}  = \tilde{f} (E(\vec{r},t)) -\frac{E}{{\tau}_c} + \left[\tilde{ \alpha_1} \frac {\partial f}{\partial E}\right] \frac {\partial E}{\partial x} 
\end{equation}
where ${\tau}_c$ is cavity lifetime.Note the similar equation appears also in acoustics \cite{Hohenberg:1993}. 
When  $  {\alpha_3} = 0$  the  Burgers-like  equation with nonlinear source $ f $ follows:  
\begin{eqnarray}
\label{burgers}
\ E_{n+1}(\vec{r} ) = f (E_{n}(\vec{r}))  + \left[\alpha_1 \frac {\partial f}{\partial E}\right] \frac {\partial E_n}{\partial x} + 
	\left[\tilde{\alpha_2}\frac {\partial f}{\partial E}\right]\frac {\partial^2 E_n}{\partial x^2}
\end{eqnarray}
or for continuous time $t$:
\begin{eqnarray}
\label{burgers1}
\frac {\partial E(\vec{r},t)}{\partial t}  = \tilde{f} (E(\vec{r},t)) -\frac{E}{{\tau}_c} +  \left[\tilde{ \alpha_1} \frac {\partial f}{\partial E}\right] \frac {\partial E}{\partial x} + 
	\left[\tilde{\alpha_2}\frac {\partial f}{\partial E}\right]\frac {\partial^2 E}{\partial x^2}
\end{eqnarray}

When     $  {\alpha_2} = 0$  the  analog of  the  Korteveg-De-Vries  equation (again with nonlinear source$ f $ ) occurs:  
\begin{equation}
\label{kdv}
\ E_{n+1}(\vec{r} ) = f (E_{n}(\vec{r}))  +  \left[\alpha_1 \frac {\partial f}{\partial E}\right] \frac {\partial E_n}{\partial x}
	 + \left[ {\alpha_3}\frac {\partial f}{\partial E}\right]\frac {\partial^3 E_n}{\partial x^3}
\end{equation}
or for continuous time $t$:
\begin{equation}
\label{kdv1}
\frac {\partial E(\vec{r},t)}{\partial t}  = \tilde{f} (E(\vec{r},t)) -\frac{E}{{\tau}_c}+   \left[\tilde{ \alpha_1} \frac {\partial f}{\partial E}\right] \frac {\partial E}{\partial x}
	 +  \left[\tilde{\alpha_3}\frac {\partial f}{\partial E}\right]\frac {\partial^3 E}{\partial x^3}
\end{equation}
\section{Low gain limit. }

Now turn our attention to the low gain limit:  nonlinearity and dispersion are of the same order  $  {\theta} $ . 
We are applying here the same "formal" trick as in \cite{Okulov:1988}  using the following identities: 
\begin{equation}
\label{approx}
\ K(x-x')= \delta(x-x') + \underbrace{\left\{K(x-x') - \delta(x-x')\right\}}
\end{equation}
and 
\begin{equation}
\label{approx1}
\ f (E_{n}(x))  = E_{n}(x) +   \underbrace{\left\{f(E_{n}(x)) - E_{n}(x)\right\}}
\end{equation}
In  (\ref{approx}) and (\ref{approx1}) the small terms underbraced by figure brackets, are responsible for nonlinearity and 
dispersion. They are assumed to be of the order  $  {\theta} $ of smallness .
Hence we have another secondary master equation, by substitution of (\ref{approx}) and (\ref{approx1}) to (\ref{nonlmap}) : 
\begin{eqnarray}
\label{master}
\ E_{n+1}(\vec{r} ) &=& f (E_{n}(\vec{r})) - E_{n}(\vec{r}) +  \hat{K}  E_{n}(\vec{r})+\nonumber \\
&+&  \underbrace {\left\{ \hat{K} E_{n}(x) - E_{n}(x)\right\} {\left\{f(E_{n}(x)) - E_{n}(x)\right\}}}  
\end{eqnarray}
Because the last term in (\ref{master}) is of the order $  {\theta}^2 $, it could be omitted and we have the secondary 
master equation in the following form:
\begin{equation}
\label{master1}
\ E_{n+1}(\vec{r} ) = f (E_{n}(\vec{r})) - E_{n}(\vec{r}) +  \hat{K}  E_{n}(\vec{r})
\end{equation}
and for continuous time: 
\begin{equation}
\label{master2}
\frac {\partial E(\vec{r},t)}{\partial t}  = \tilde{f} (E(\vec{r},t)) -\frac{E}{{\tau}_c} +  \hat{\tilde{K} } (E(\vec{r},t)) 
\end{equation}
Applying the same series expansion to the convolution inside (\ref{master1}) we have: 
\begin{equation}
\label{master3}
\ E_{n+1}(\vec{r} ) = f (E_{n}(\vec{r})) - E_{n}(\vec{r}) + \sum^\infty_{m=1}\alpha_m \frac {\partial^m E_n}{\partial x^m}
\end{equation}	
where  $ \alpha_m$ 
are again the moments (\ref{moment}) of the kernel $  {K}$. Consequently 
for continuous time one may write: 
\begin{equation}
\label{master4}
\frac {\partial E(\vec{r},t)}{\partial t}  = \tilde{f} (E(\vec{r},t)) -\frac{E}{{\tau}_c} +  \sum^\infty_{m=1}\tilde{\alpha_m} \frac {\partial^m E}{\partial x^m}
\end{equation}

As a further approximation one can get readily the several familiar 
dynamic equations, depending on relative magnitude of real and imaginary parts 
of the moments of kernel and  relative strength of the nonlinearity. For example 
for all $ {\alpha_m}=0 $ except for  $  {m=2}$ and  $  {\alpha_2}$ - real (pure confocal cavity),  
Kolmogorov-Petrovsky-Piskounov equation follows (KPP) \cite{Okulov:1988,Okulov:2000}: 
\begin{equation}
\label{kpp}
\ E_{n+1}(x) = f (E_n(x)) - E_n(x) + \alpha_2 \frac {\partial^2 E_n}{\partial x^2}
\end{equation}	
For continuous time this equation has the following form:
\begin{equation}
\label{kpp1}
\frac {\partial E(\vec{r},t)}{\partial t}  = \tilde{f} (E(\vec{r},t)) -\frac{E}{{\tau}_c} +  \tilde{\alpha_2} \frac {\partial^2 E}{\partial x^2}
\end{equation}	
Next, for the $ {\alpha_1}=0 $ ,   $ {\alpha_2} $  - purely imaginary (plane-parallel Fabry-Perot cavity),  
and  the $ {f (E_n(x))= i k{n_2}d E_n |E|^2 }$  nonlinear Shrodinger equation  (NLS) occurs \cite{Newell:1990,Okulov:1988}: 
\begin{equation}
\label{nls}
\ E_{n+1}(x ) = i Im ( \alpha_2 ) \frac {\partial^2 E_n}{\partial x^2  }- E_{n}+ i k{n_2} d E_n  |E_n|^2
\end{equation}	
For continuous time we have: 
\begin{equation}
\label{nls1}
\frac {\partial E(\vec{r},t)}{\partial t}  =  i Im (\tilde{\alpha_2}) \frac {\partial^2 E}{\partial x^2}-\frac{E}{{\tau}_c} + i c k{n_2} E |E|^2
\end{equation}	
Furthermore, for all moments of kernel $K$ $ {\alpha_m}=0 $ except for  $  {m=2}$  ,  
and  $ {\alpha_2} $ - arbitrarily nonzero complex number ( Fabry-Perot cavity with curved mirrors)  and  
\begin{equation}
\label{semicond}
\ f (E_n(x))= G d E_n  \left[ 1 + (Re {\beta}) E_n  |E_n|^2 + i (Im {\beta}) E_n  |E_n|^2\right]  
\end{equation}	
(semiconductor or solid-state gain medium) the 
Ginzburg-Landau equation (GLE) is valid \cite{Okulov:1988}:  
\begin{eqnarray}
\label{gle}
\ E_{n+1}(x ) =  Re( \alpha_2 ) \frac {\partial^2 E_n}{\partial x^2  }+ i Im ( \alpha_2 ) \frac {\partial^2 E_n}{\partial x^2  }- E_n(x) + \nonumber \\
+ {G d E_n { \left[1 + (Re{\beta})E_n |E_n|^2 + i (Im{\beta})E_n |E_n|^2 \right]}}
\end{eqnarray}	
For continuous time we have: 
\begin{equation}
\label{gle1}
\frac {\partial E(\vec{r},t)}{\partial t}  =  Re (\tilde{\alpha_2}) \frac{\partial^2 E}{\partial x^2  }-\frac{E}{{\tau}_c}+
i Im (  \tilde{\alpha_2}) \frac {\partial^2 E}{\partial x^2  } +  \tilde{G} E +  (Re\tilde{ {\beta}})E |E|^2 + i  (Im\tilde {\beta} )E |E|^2 
\end{equation}	
In a final step of our consideration, for the only nonzero  $ {\alpha_2} $  and   $ {\alpha_4} $  - arbitrarily complex numbers 
 ( corresponding to Fabry-Perot cavity with curved mirrors)  and  
\begin{equation}
\label{semicond1}
\ f (E_n(x))= G E_n + {\beta} E_n  |E_n|^2 +  {\gamma} E_n  |E_n|^4
\end{equation}	
( semiconductor or solid-state gain medium in deep 
saturation ) the Swift-Hohenberg equation  (SHE) tooks  place \cite{Okulov:2002,Hohenberg:1993}:
\begin{eqnarray}
\label{she}
\ E_{n+1}(x ) = \alpha_2  \frac {\partial^2 E_n}{\partial x^2  }+
  \alpha_4  \frac {\partial^4 E_n}{\partial x^4  }- E_n(x)+ G E_n +  {\beta}E_n |E_n|^2 + {\gamma}E_n |E_n|^4
\end{eqnarray}
And at last we have for continuous time: 
\begin{eqnarray}
\label{she1}
\frac {\partial E(\vec{r},t)}{\partial t}  =  \tilde{\alpha_2}\frac {\partial^2 E}{\partial x^2  }+
\tilde{  \alpha_4}  \frac {\partial^4 E}{\partial x^4  }-\frac{E}{{\tau}_c}+  \tilde{G} E +  \tilde{ {\beta}}E |E|^2 + \tilde{ {\gamma}}E |E|^4
\end{eqnarray}
\section{Conclusion.}

	We have just saw that possible dynamical regimes of the electromagnetic field in microlaser cavity is much 
wider than those 
simulated by parabolic or Swift-Hohenberg equations scenarious. The boundary conditions itself affect the form of 
the master equation, hence they affect the observed laser dynamics.

It is noteworthy to keep in mind that each of the evolution equations derived above have 
its own peculiar behaviour. Some of these equations,  like KPP (\ref{kpp})or  NLSE (\ref{nls}) 
have exact solutions in the form of kinks (switching waves ) or solitons . As for the other equations,
 like GLE (\ref{gle})  or SHE (\ref{she}), their exact solutions are not known, although there exists a set of 
established analytical, semianalytical and numerical results, showing which dynamics is to be expected in the 
given range of parameters. Moreover  
by means of choosing  the special resonator configuration, setting the appropriate gain-losses distribution  
$ {G} $ and  nonlinearity
 it is possible to simulate the spatiotemporal dynamics of a given dynamical equation , say  GLE
  (\ref{gle}) or  SHE equations (\ref{she}). The initial conditions will evolve along the phase trajectories of the 
evolution equation to be investigated. This provides a possibility of the purely optical analog modeling of  partial 
differential equations (\ref{shock}, \ref{kdv}, \ref{gle}, \ref{kpp}, \ref{nls}, \ref{gle}, \ref{she}) extracted above from nonlocal maps
(coupled map lattices). 
On the other hand because this approach is going to be useful also in condensed matter physics, for example as a model of 
granular materials  \cite{Ott:2001}: the results presented here provide the definite information concerning the expected 
dynamical regimes 
of the pattern formation and properties of this materials. 

\section{Acknowledgements.}
The author acknowleges Department of Physics of the University of Coimbra for hospitality.
This work was partially supported by NATO ICCTI grant.

\baselineskip 12pt plus 2pt minus 2pt


\begin{thebibliography}{99}

\bibitem{Oraevsky:1964} A.Z.Grasyuk, A.N.Oraevsky., {\it  Radiotekhnika i Electronika} {\bf 9}, 524 (1964).
\bibitem{Lorenz:1963} E.N.Lorenz., {\it  J.Atmos.Sci.} {\bf 20}, 130  (1963).
\bibitem{Newell:1990} A.C.Newell, J.V.Moloney., {\it Physica D}  {\bf 44}, 1-37 (1990).
\bibitem{Okulov:1988} A.Yu.Okulov, A.N.Oraevsky., {\it Proceedings P.N.Lebedev Physics Institute}(in Russian),
 N.G.Basov ed., Nauka, Moscow. {\bf 187}, 202-222 (1988). Library of Congress Control Number: 88174540 ( http://www.loc.gov).
\bibitem{Lugiato:1991} M.Brambilla, F.Battipede, L.A.Lugiato, V.Penna, F.Prati,C. Tamm, and C.O.Weiss., {\it Phys.Rev.A} {\bf 43}, 5090 (1991). 
\bibitem{Staliunas:1993} K.Staliunas., {\it Phys.Rev. A} {\bf 48}, 1573 (1993). 
K.Staliunas, C.O.Weiss., {\it JOSA B} {\bf 12}, 1142  (1995).
\bibitem{Rosanov:2003} S.V.Fedorov, N.N.Rozanov, A.N.Shatsev, N.A.Veretenov, A.G.Vladimirov., {\it IEEE-QE} {\bf v.39}, N.2,197 (2003). 
\bibitem{Lugiato:1998} L.Spinelli, G.Tissoni, M.Brambilla, F.Prati, L.A.Lugiato., {\it Phys.Rev. A} {\bf 58}, 2542 (1998). 
\bibitem{Firth:1998} G.K.Harkness,R.Martin,  A.J.Scroggie,G.L.Oppo,  W.J.Firth., {\it Phys.Rev. A} {\bf 58}, 2577 (1998). 
\bibitem{Okulov:2000}A.Yu.Okulov.,{\it Optics and Spectroscopy } {\bf 89}, 145  (2000).
\bibitem{Barabashi:2002}R.Albert,A.L.Barabashi., {\it Rev. Mod.Phys} {\bf 74}, 47 (2002). 
\bibitem{Weinstein:1969} L.A.Weinstein., {\it "Open resonators and open waveguides". Golem Press, Colorado}(1969). 
\bibitem{Okulov:1990}A.Yu.Okulov.,{\it JOSA B} {\bf 7}, 1045  (1990).
\bibitem{Chen:2001} Y.F.Chen, Y.P.Lan., {\it Phys.Rev.A} {\bf 64}, 063807 (2001). 
\bibitem{Okulov:2002} A.Yu.Okulov., {\it P.N.Lebedev Physical Institute, Preprint N24} (2002). 
\bibitem{Staliunas:1998} K.Staliunas, V.B.Taranenko, G.Slekys, R.Viselga, C.O.Weiss., {\it Phys.Rev. A} {\bf 57}, 599 (1998). 
\bibitem{Pikovsky:1986}S.P.Kuznetsov, A.S.Pikovsky., {\it Physica D}  {\bf 19}, 384 (1986).
\bibitem{Ott:2001}S.C.Venkataramani, E.Ott., {\it Phys.Rev. E} {\bf 63}, 046202 (2001).
\bibitem{Suchkov:1965}A.F.Suchkov., {\it JETP} {\bf 22}, 1026 (1966).
\bibitem{Hohenberg:1993}M.Cross,P.C.Hohenberg., {\it Rev. Mod.Phys} {\bf 65}, 851 (1993).
\end{thebibliography}
\end{document}